\documentclass[12pt]{article}
\usepackage{a4wide}
\usepackage{latexsym}
\usepackage{cite}
\usepackage{axodraw}

\usepackage{pslatex}
\usepackage[latin1]{inputenc}
\usepackage[T1]{fontenc}

\def\bq{\begin{eqnarray}}
\def\eq{\end{eqnarray}}
\def\l{\langle}
\def\r{\rangle} 
\def\eps{\varepsilon}
\def\v{\verb}

\begin{document}

\thispagestyle{empty}

\begin{flushright}
  MZ-TH/05-22 \\
\end{flushright}

\vspace{1.5cm}

\begin{center}
  {\Large\bf 
   Automated computation of spin- and colour-correlated Born matrix elements\\
  }
  \vspace{1cm}
  {\large 
    Stefan Weinzierl\\
  \vspace{1cm}
      {\small \em Institut f{\"u}r Physik, Universit{\"a}t Mainz,}\\
      {\small \em D - 55099 Mainz, Germany}\\
  } 
\end{center}

\vspace{2cm}

\begin{abstract}\noindent
  {
   I report on an implementation of an algorithm for the automated numerical calculation
   of spin- and colour-correlated Born matrix elements in QCD. These spin- and colour-correlated matrix elements are
   needed for NLO calculations in combination with the subtraction method.
   Both massless and massive quarks are considered.
   There are no restrictions on the number of external particles.
   As a trivial sub-case, the algorithm also applies to Born matrix elements without any correlations.
   These are sufficient for leading order calculations. 
   }
\end{abstract}

\vspace*{\fill}

\newpage

\section{Introduction}
\label{sect:intro}

QCD processes will constitute the bulk of events at the LHC. These processes provide
information on the strong interaction and form quite often important background for
searches of new physics.
An accurate description of jet physics is therefore mandatory.
Although jet observables can rather easily be modelled at leading order (LO)
in perturbation
theory \cite{Berends:1987me,Berends:1989ie,Berends:1990ax,Caravaglios:1995cd,Caravaglios:1998yr,Draggiotis:1998gr,Draggiotis:2002hm,Stelzer:1994ta,Pukhov:1999gg,Yuasa:1999rg,Krauss:2001iv},
this description suffers several drawbacks.
A leading order calculation depends strongly on the renormalization scale
and can therefore give only an order-of-magnitude-estimate on absolute rates.
Secondly, at leading order a jet is modelled by a single parton. This is a very crude
approximation and oversimplifies inter- and intra-jet correlations.
The situation is improved by including higher-order corrections in perturbation theory.

At present, there are many next-to-leading order (NLO) calculation for $2 \rightarrow 2$ processes at
hadron colliders, but only a few for $2 \rightarrow 3$ processes.
Fully differential numerical programs exist for example for
$p p \rightarrow \mbox{3 jets}$ \cite{Kilgore:1996sq,Nagy:2001fj,Nagy:2003tz},
$p p \rightarrow V + \mbox{2 jets}$ \cite{Campbell:2002tg},
$p p \rightarrow t \bar{t} H$ \cite{Beenakker:2002nc,Dawson:2003zu} and
$p p \rightarrow H + \mbox{2 jets}$ \cite{DelDuca:2001eu,DelDuca:2001fn}.

It is desirable to have NLO calculations for $2 \rightarrow n$ processes in hadron-hadron
collisions with $n$ in the range of $n=3,4,...,6,7$.
QCD processes like $p p \rightarrow n \; \mbox{jets}$ form often important backgrounds for the searches of
signals of new physics.
However, the complexity of the calculation increases
with the number of final state particles.
To overcome the computational limitation, there have been in the past years
several proposals for the automated computation
of next-to-leading order observables
\cite{Soper:1998ye,Soper:1999xk,Passarino:2001wv,Ferroglia:2002mz,Nagy:2003qn,Denner:2002ii,Dittmaier:2003bc,Giele:2004iy,Ellis:2005zh,delAguila:2004nf,Pittau:2004bc,vanHameren:2005ed,Binoth:2002xh,Binoth:2005ff,vanHameren:2004wr}.
These publications focussed mainly on the automated computation of loop integrals.
Equally important is the computation of the real emission contribution.
It is well known that in general an NLO observable will receive contributions from the virtual corrections
and the real emission part.
Taken separately, each of the two contributions is divergent due to the presence of infrared singularities.
Only the sum of the two is finite.
There are several general method available to handle this problem, like 
the phase-space slicing method
\cite{Giele:1992vf,Giele:1993dj,Keller:1998tf}
or the subtraction method
\cite{Frixione:1996ms,Catani:1997vz,Catani:1997vzerr,Dittmaier:1999mb,Phaf:2001gc,Catani:2002hc}.
In this paper I will focus on the dipole subtraction method 
\cite{Catani:1997vz,Catani:1997vzerr,Dittmaier:1999mb,Phaf:2001gc,Catani:2002hc}.
The dipole subtraction method requires the calculation of spin- and colour-correlated Born matrix elements.
In this paper I describe a method for the automated calculation of these quantities.
While the kinematical part of the matrix elements is calculated numerically, colour-correlation matrices
are calculated symbolically at the initialisation phase of the program.
The C++ library ``GiNaC'' \cite{Bauer:2000cp}
allows to mix numerical and symbolical code in a single program.
The program uses standard techniques like spinor methods
\cite{Berends:1981rb,DeCausmaecker:1982bg,Gunion:1985vc,Xu:1987xb,Gastmans:1990xh}
and colour decomposition
\cite{Cvitanovic:1980bu,Berends:1987cv,Mangano:1987xk,Kosower:1987ic,Bern:1990ux,DelDuca:1999rs,Maltoni:2002mq}.
The program computes helicity amplitudes, which are decomposed into colour factors and partial amplitudes.
The partial amplitudes are computed with the help of Berends-Giele type recurrence relations
\cite{Berends:1987me,Kosower:1989xy}.
It should be noted that recently interesting new methods emerged for the computation of partial amplitudes
\cite{Cachazo:2004kj,Britto:2004ap,Bena:2004ry,Schwinn:2005pi}.

This paper is organised as follows:
In the following section I present the general setup for the dipole subtraction method
and review a few basic tools for the calculation of QCD amplitudes.
Sect. \ref{sect:method} describes the algorithm for the calculation of colour-correlated Born 
matrix elements.
The numerical implementation is discussed
in sect. \ref{sect:num}.
Finally, sect. \ref{sect:concl} contains the conclusions and an outlook.
In an appendix I summarise the colour-ordered Feynman rules and the colour-correlation operators.
Furthermore, I give some technical details on the implementation into a C++ program.


\section{General setup and basic tools}
\label{sect:basic_tools}

\subsection{The dipole formalism}

The starting point for the calculation of an infrared safe observable $O$ in
hadron-hadron collisions is the following formula:
\bq
\l O \r & = & \int dx_1 f(x_1) \int dx_2 f(x_2) \frac{1}{2 K(\hat{s})}
             \frac{1}{\left( 2 J_1+1 \right)}
             \frac{1}{\left( 2 J_2+1 \right)} 
             \frac{1}{n_1 n_2}
 \\
 & &
             \int d\phi_{n}\left(p_1,p_2;p_3,...,p_{n+2}\right)
             O\left(p_1,...,p_{n+2}\right)
             \left| {\cal A}_{n+2} \right|^2.
\eq
This equation gives the contribution from the $n$-parton final state.
The two incoming particles are labelled $p_1$ and $p_2$, while $p_3$ to $p_{n+2}$ denote
the final state particles.
$f(x)$ gives the probability of finding a parton $a$ with momentum fraction $x$ inside
the parent hadron $h$.
A sum over all possible partons $a$ is understood implicitly.
$2K(s)$ is the flux factor,
$1/(2J_1+1)$ and $1/(2J_2+1)$ correspond
to an averaging over the initial helicities and
$n_1$ and $n_2$ are the number of colour degrees  of the initial state particles.
$d\phi_n$ is the phase space measure for $n$ final state particles, including (if appropriate) the identical particle factors.
The matrix element $| {\cal A}_{n+2} |^2$ is calculated perturbatively.

At NLO one has the following contributions:
\bq
\l O \r^{NLO} & = & 
 \int\limits_{n+1} O_{n+1} d\sigma^R + \int\limits_n O_n d\sigma^V 
 + \int\limits_n O_n d\sigma^C.
\eq
Here I used a rather condensed notation. $d\sigma^R$ denotes the real emission contribution,
whose matrix element is given by the square of the Born amplitudes with $(n+3)$ partons
$| {\cal A}^{(0)}_{n+3} |^2$.
$d\sigma^V$ gives the virtual contribution, whose matrix element is given by the interference term
of the one-loop amplitude ${\cal A}^{(1)}_{n+2}$ with $(n+2)$ partons with the corresponding
Born amplitude ${\cal A}^{(0)}_{n+2}$.
$d\sigma^C$ denotes a collinear subtraction term, which subtracts the initial-state collinear
singularities.
Taken separately, the individual contributions are divergent and only their sum is finite.
In order to render the individual contributions finite, such that the phase space integrations
can be performed by Monte Carlo methods, one adds and subtracts a suitable chosen piece
\cite{Catani:1997vz,Catani:1997vzerr,Dittmaier:1999mb,Phaf:2001gc,Catani:2002hc}:
\bq
\l O \r^{NLO} & = & 
 \int\limits_{n+1} \left( O_{n+1} d\sigma^R - O_n d\sigma^A \right)
 + \int\limits_n \left( O_n d\sigma^V + O_n d\sigma^C + O_n \int\limits_1 d\sigma^A \right).
\eq
The matrix element corresponding to the approximation term $d\sigma^A$ is given as a sum over 
dipoles:
\bq
 \sum\limits_{pairs\; i,j} \;\;\; \sum\limits_{k \neq i,j} {\cal D}_{ij,k}.
\eq
Each dipole contribution has the following form:
\bq
{\cal D}_{ij,k} 
& = & 
- \frac{1}{2 p_i \cdot p_j}
{\cal A}_{n+2}^{(0)\;\ast}\left( p_1, ..., \tilde{p}_{(ij)},...,\tilde{p}_k,...\right)
\frac{{\bf T}_k \cdot {\bf T}_{ij}}{{\bf T}^2_{ij}} V_{ij,k} 
{\cal A}_{n+2}^{(0)}\left( p_1, ..., \tilde{p}_{(ij)},...,\tilde{p}_k,...\right).
\eq
Here ${\bf T}_i$ denotes the colour charge operator \cite{Catani:1997vz} for parton $i$ and
$V_{ij,k}$ is a matrix in the spin space of the emitter parton $(ij)$.
Explicit formulae for the expressions $V_{ij,k}$ can be found in the literature 
\cite{Catani:1997vz,Catani:1997vzerr,Dittmaier:1999mb,Phaf:2001gc,Catani:2002hc}
and are not repeated here.
In the numerical program both the dipole terms for massless and massive partons are implemented.

In general, the operators ${\bf T}_i$ lead to colour correlations, while the $V_{ij,k}$'s lead
to spin correlations.
The colour charge operators ${\bf T}_i$ for a quark, gluon and antiquark in the final state are
\bq
\label{colour_charge_operator_final}
\mbox{quark :} & & 
 {\cal A}^\ast\left(  ... q_i ... \right) \left( T_{ij}^a \right) {\cal A}\left(  ... q_j ... \right), \nonumber \\
\mbox{gluon :} & & 
 {\cal A}^\ast\left(  ... g^c ... \right) \left( i f^{cab} \right) {\cal A}\left(  ... g^b ... \right), \nonumber \\
\mbox{antiquark :} & & 
 {\cal A}^\ast\left(  ... \bar{q}_i ... \right) \left( - T_{ji}^a \right) {\cal A}\left(  ... \bar{q}_j ... \right).
\eq
The corresponding colour charge operators for a quark, gluon and antiquark in the initial state are
\bq
\label{colour_charge_operator_initial}
\mbox{quark :} & & 
 {\cal A}^\ast\left(  ... \bar{q}_i ... \right) \left( - T_{ji}^a \right) {\cal A}\left(  ... \bar{q}_j ... \right), \nonumber \\
\mbox{gluon :} & & 
 {\cal A}^\ast\left(  ... g^c ... \right) \left( i f^{cab} \right) {\cal A}\left(  ... g^b ... \right), \nonumber \\
\mbox{antiquark :} & & 
 {\cal A}^\ast\left(  ... q_i ... \right) \left( T_{ij}^a \right) {\cal A}\left(  ... q_j ... \right). 
\eq
In the amplitude an incoming quark is denoted as an outgoing antiquark and vice versa.

The subtraction term can be integrated over the unresolved one-parton phase space.
Due to this integration, all spin-correlations average out, but colour correlations still remain.
In a compact notation, the result of this integration is often written as
\bq
 d\sigma^C + \int\limits_1 d\sigma^A  
 & = & {\bf I} \otimes d\sigma^B + {\bf K} \otimes d\sigma^B + {\bf P} \otimes d\sigma^B.
\eq
The notation $\otimes$ indicates that colour correlation still remain.
The term ${\bf I} \otimes d\sigma^B$ lives on the phase space of the $n$-parton configuration and has the appropriate
singularity structure to cancel the infrared divergences coming from the one-loop amplitude.
Therefore $d\sigma^V + {\bf I} \otimes d\sigma^B$
is infrared finite.

The purpose of the paper is to set up a numerical program for the automated computation of the
terms
\bq  
\label{task1}
 \int\limits_{n+1} \left( O_{n+1} d\sigma^R - O_n d\sigma^A \right)
\eq
and 
\bq
\label{task2}
 \int\limits_n O_n \left( {\bf I} \otimes d\sigma^B + {\bf K} \otimes d\sigma^B + {\bf P} \otimes d\sigma^B \right).
\eq
This requires the computation of the matrix elements with $(n+3)$ partons with no spin- or colour-correlations (implicit in
$d\sigma^R$) as well as the computation of matrix elements with $(n+2)$ partons with spin- and colour-correlations.
The subtraction terms in eq. (\ref{task1}) involve spin and colour correlations. 
The insertion operators ${\bf I}$, ${\bf K}$ and ${\bf P}$ induce colour correlations, but no spin correlations.
One is therefore naturally lead to
the calculation of colour-ordered amplitudes in a helicity basis.
Basic techniques for such a task are reviewed in the next subsections.

\subsection{Double line notation}

In QCD one deals with quarks and gluons. Both types of partons carry information on the colour degrees of freedoms
and the kinematical degrees of freedom.
Quarks have a colour index $i$, running from $1$ to $N$ and corresponding to the fundamental representation of $SU(N)$.
The kinematical information can be represented for massless quarks by Weyl spinors
$p_A$ or $p_{\dot{B}}$, where the indices $A$ or $\dot{B}$ run from $1$ to $2$.
The corresponding information for gluons is in the conventional approach represented by a colour index $a$, running
from $1$ to $N^2-1$ and which corresponds to the adjoint representation of $SU(N)$.
The kinematical information is represented by a Lorentz index $\mu$, running from
$0$ to $3$.
It is useful, to treat quarks and gluons on the same footing. To this aim, I follow the ``double-line''-approach
\cite{'tHooft:1973jz}
and convert a gluon index to two quark indices. I do this for the colour degrees of freedom, as well as for the kinematical
parts.

In detail, this is done as follows: In Feynman diagrams one distinguishes edges and vertices.
Edges are propagators as well as polarisation vectors or spinors for external
particles.
Vertices are all interaction vertices.
For vector-like couplings one can write
\bq
V_\mu E^\mu & = & V_\mu g^{\mu\nu} E_\nu 
 = V_\mu \left( \frac{1}{2} \sigma^\mu_{A\dot{B}} \bar{\sigma}^{\nu\dot{B}A} \right) E_\nu
 = 
 \left( \frac{1}{\sqrt{2}} V_\mu \sigma^\mu_{A\dot{B}} \right)
 \left( \frac{1}{\sqrt{2}} \bar{\sigma}^{\nu\dot{B}A} E_\nu \right),
\eq
which allows us to replace a contraction over $\mu$ by two contractions over $A$ and $\dot{B}$.
One can apply the same trick to the colour algebra:
\bq
V^a E^a & = & V^a \delta^{ab} E^b = V^a \left( 2 T^a_{ij} T^b_{ji} \right) E^b
 =  \left( \sqrt{2} T^a_{ij} V^a \right) \left( \sqrt{2} T^b_{ji} E^b \right).
\eq
Again, this equation allows us to replace a contraction over an adjoint index $a$ by two contractions over
indices $i$ and $j$ in the fundamental representation.
The Feynman rules for QCD in the double line notation are listed 
in appendix \ref{appendix:feynman}.

\subsection{Colour decomposition}
In this paper I use the normalisation
\bq
 \mbox{Tr}\;T^a T^b & = & \frac{1}{2} \delta^{a b}
\eq
for the colour matrices.
Amplitudes in QCD may be decomposed into group-theoretical factors (carrying the colour structures)
multiplied by kinematic functions called partial amplitudes
\cite{Cvitanovic:1980bu,Berends:1987cv,Mangano:1987xk,Kosower:1987ic,Bern:1990ux}. 
These partial amplitudes do not contain any colour information and are gauge-invariant objects. 
\\
\\
The colour decomposition is obtained by replacing the structure constants $f^{abc}$
by
\bq
 i f^{abc} & = & 2 \left[ \mbox{Tr}\left(T^a T^b T^c\right) - \mbox{Tr}\left(T^b T^a T^c\right) \right] 
\eq
which follows from $ \left[ T^a, T^b \right] = i f^{abc} T^c$.
The resulting traces and strings of colour matrices can be further simplified with
the help of the Fierz identity :
\bq
 T^a_{ij} T^a_{kl} & = &  \frac{1}{2} \left( \delta_{il} \delta_{jk}
                         - \frac{1}{N} \delta_{ij} \delta_{kl} \right).
\eq
In the pure gluonic case tree level amplitudes with $n$ external gluons may be written in 
the form
\bq
\label{colour_decomp_pure_gluon}
{\cal A}_{n}(1,2,...,n) & = & \left(\frac{g}{\sqrt{2}}\right)^{n-2} \sum\limits_{\sigma \in S_{n}/Z_{n}} 
 \delta_{i_{\sigma_1} j_{\sigma_2}} \delta_{i_{\sigma_2} j_{\sigma_3}} 
 ... \delta_{i_{\sigma_n} j_{\sigma_1}}  
 A_{n}\left( \sigma_1, ..., \sigma_n \right),
\eq
where the sum is over all non-cyclic permutations of the external gluon legs.
The quantities $A_n(\sigma_1,...,\sigma_n)$, called the partial amplitudes, contain the 
kinematic information.
They are colour-ordered, e.g. only diagrams with a particular cyclic ordering of the gluon
s contribute.
The choice of the basis for the colour structures is not unique, and several proposals
for bases can be found in the literature \cite{DelDuca:1999rs,Maltoni:2002mq}.
Here I use the ``colour-flow decomposition'' \cite{Maltoni:2002mq}.
As a further example I give the
the colour decomposition for a tree amplitude with a pair of quarks:
\bq
{\cal A}_{n+2}(q,1,2,...,n,\bar{q}) 
& = & \left(\frac{g}{\sqrt{2}}\right)^{n} 
 \sum\limits_{S_n} 
 \delta_{i_q j_{\sigma_1}} \delta_{i_{\sigma_1} j_{\sigma_2}} 
 ... \delta_{i_{\sigma_n} j_{\bar{q}}} 
A_{n+2}(q,\sigma_1,\sigma_2,...,\sigma_n,\bar{q}). 
\eq
where the sum is over all permutations of the gluon legs. 
In squaring these amplitudes a colour projector
\bq
 \delta_{\bar{i} i} \delta_{j \bar{j}} - \frac{1}{N} \delta_{\bar{i} \bar{j} } \delta_{j i}
\eq
has to applied to each gluon.

While the colour structure of the examples quoted above is rather simple, the colour decomposition
can be become rather involved for amplitudes with many pairs of quarks.
A systematic algorithm for the colour decomposition and the diagrams contributing to a single colour
structure is given in sect. \ref{sect:method}.

\subsection{Spinor techniques}

For the calculation of 
helicity amplitudes \cite{Berends:1981rb,DeCausmaecker:1982bg,Gunion:1985vc,Xu:1987xb,Gastmans:1990xh}
one chooses for the 
spinors corresponding to external massless quarks two-component Weyl spinors.
Two notations for Weyl spinors are commonly used in the literature.
The relation between the bra-ket notation and the notation using dotted and undotted indices
is as follows:
\bq
|p+\rangle = p_B,          & & \langle p+| = p_{\dot{A}}, \\
|p-\rangle = p^{\dot{B}},  & & \langle p-| = p^A. 
\eq
Spinor products are denoted as follows:
\bq 
 \l p q \r = \l p- | q+ \r = p^A q_A,
 & &
 [ q p ] = \l q+ | p- \r = q_{\dot{A}} p^{\dot{A}}.
\eq
For the polarisation vectors of the external gluons one uses
\bq
\label{gluon_pol_onshell}
\eps_{\mu}^{+}(k,q) = \frac{\langle q-|\gamma_{\mu}|k-\rangle}{\sqrt{2} \langle q- | k + \rangle},
 & &
\eps_{\mu}^{-}(k,q) = \frac{\langle q+|\gamma_{\mu}|k+\rangle}{\sqrt{2} \langle k + | q - \rangle},
\eq
where $k$ is the momentum of the gluon and $q$ is an arbitrary light-like reference momentum.
In the ``double-line'' notation this becomes
\bq
\eps^{\dot{A}B}_+(k,q) =  
 \frac{1}{\l q k \r} \; k^{\dot{A}} q^B,
 & &
\eps^{\dot{A}B}_-(k,q) =  
 \frac{1}{[ k q ]} \; q^{\dot{A}} k^B.
\eq
For spinors corresponding to massive quarks the formulae from ref. \cite{vanderHeide:2000fx} are used.

\subsection{Recurrence relations}
\label{subsect:recurrence}

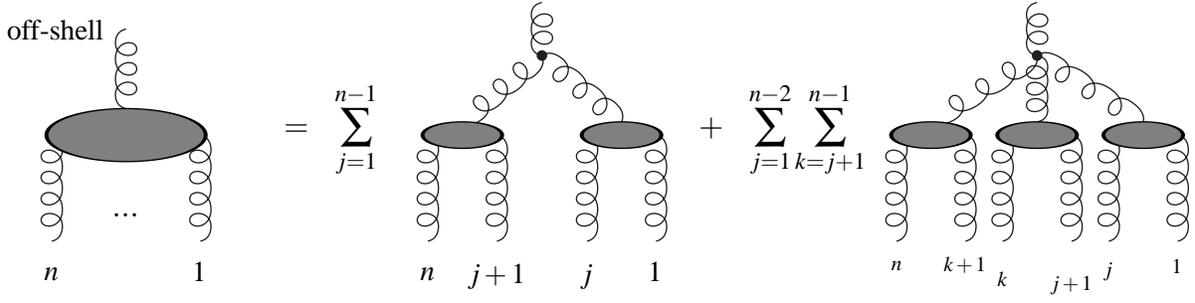
\begin{figure}
\begin{center}
\vspace*{-20mm}
\bq
\begin{picture}(100,100)(0,50)
 \Gluon(50,60)(50,90){4}{3}
 \Gluon(78,50)(78,10){4}{4}
 \Gluon(22,50)(22,10){4}{4}
 \GOval(50,50)(10,30)(0){0.5}
 \Text(50,20)[c]{$...$}
 \Text(78,-5)[b]{\small $1$}
 \Text(22,-5)[b]{\small $n$}
 \Text(42,90)[r]{\small off-shell}
\end{picture}
 & = & 
 \sum\limits_{j=1}^{n-1}
 \;\;\;\;
\begin{picture}(100,100)(0,50)
 \Vertex(50,80){2}
 \Gluon(50,80)(50,100){4}{2}
 \Gluon(50,80)(80,55){4}{3}
 \Gluon(50,80)(20,55){4}{3}
 \Gluon(33,50)(33,10){4}{4}
 \Gluon(7,50)(7,10){4}{4}
 \Gluon(93,50)(93,10){4}{4}
 \Gluon(67,50)(67,10){4}{4}
 \GOval(20,50)(5,15)(0){0.5}
 \GOval(80,50)(5,15)(0){0.5}
 \Text(93,-5)[b]{\small $1$}
 \Text(67,-8)[b]{\small $j$}
 \Text(33,-8)[b]{\small $j+1$}
 \Text(7,-5)[b]{\small $n$}
\end{picture}
 \;\;\;
+
 \;\;\;
 \sum\limits_{j=1}^{n-2}
 \sum\limits_{k=j+1}^{n-1}
 \;\;\;\;\;
\begin{picture}(100,100)(0,50)
 \Vertex(50,80){2}
 \Gluon(50,80)(50,100){4}{2}
 \Gluon(50,80)(90,55){4}{3}
 \Gluon(50,80)(10,55){4}{3}
 \Gluon(50,80)(50,55){4}{3}
 \Gluon(103,50)(103,10){4}{4}
 \Gluon(77,50)(77,10){4}{4}
 \Gluon(63,50)(63,10){4}{4}
 \Gluon(37,50)(37,10){4}{4}
 \Gluon(23,50)(23,10){4}{4}
 \Gluon(-3,50)(-3,10){4}{4}
 \GOval(10,50)(5,15)(0){0.5}
 \GOval(50,50)(5,15)(0){0.5}
 \GOval(90,50)(5,15)(0){0.5}
 \Text(103,-2)[b]{\scriptsize $1$}
 \Text(77,-5)[b]{\scriptsize $j$}
 \Text(63,-10)[b]{\scriptsize $j+1$}
 \Text(37,-7)[b]{\scriptsize $k$}
 \Text(23,-2)[b]{\scriptsize $k+1$}
 \Text(-3,-1)[b]{\scriptsize $n$}
\end{picture}
\nonumber 
\eq
\vspace*{+10mm}
\caption{\label{fig:recurrencerelation} The recurrence relation for the gluon current.
An off-shell current with $n$ legs can be computed recursively from off-shell currents with
fewer legs.}
\end{center}
\end{figure}
Recursive techniques \cite{Berends:1987me,Kosower:1989xy}
build partial amplitudes from smaller building blocks, usually
called colour-ordered off-shell currents.
Off-shell currents are objects with $n$ on-shell leg and one additional leg off-shell.
Momentum conservation is satisfied. It should be noted that
off-shell currents are not gauge-invariant objects.
Recurrence relations relate off-shell currents with $n$ legs 
to off-shell currents with fewer legs.
\\
\\
For the pure gluon current $J_{n}^{\dot{A}B}$, the recurrence relation reads
\bq
\label{recurrence_pure_gluon}
\lefteqn{
J_{n}^{\dot{A}B}\left(p_1^\pm, ..., p_{n}^\pm; q_1, ..., q_{n} \right)
 = }
\nonumber \\
 & &
 \sum\limits_{j=1}^{n-1}
  J_{j}^{\dot{C}D}\left(p_1^\pm, ..., p_{j}^\pm; q_1, ..., q_{j} \right)
  J_{n-j}^{\dot{E}F}\left(p_{j+1}^\pm, ..., p_{n}^\pm; q_{j+1}, ..., q_{n} \right)
  \nonumber \\
 & &
  \times
  V_{D\dot{C}F\dot{E}H\dot{G}}(p_{1,j},p_{j+1,n})
  P^{\dot{G}H\dot{A}B}(p_{1,n})
 \nonumber \\
 & &
 +
 \sum\limits_{j=1}^{n-2}
  \sum\limits_{k=j+1}^{n-1}
  J_{j}^{\dot{C}D}\left(p_1^\pm, ..., p_{j}^\pm; q_1, ..., q_{j} \right)
  J_{k-j}^{\dot{E}F}\left(p_{j+1}^\pm, ..., p_{k}^\pm; q_{j+1}, ..., q_{k} \right)
 \nonumber \\
 & &
  \times
  J_{n-k}^{\dot{G}H}\left(p_{k+1}^\pm, ..., p_{n}^\pm; q_{k+1}, ..., q_{n} \right)
  V_{D\dot{C}F\dot{E}H\dot{G}J\dot{I}}
  P^{\dot{I}J\dot{A}B}(p_{1,n})
\eq
This relation is pictorially shown in fig. \ref{fig:recurrencerelation}.
In this formula, the $q_i$'s are the reference momenta for the external gluons,
$P^{\dot{C}D\dot{A}B}(k)$ is the expression for the gluon propagator and
$V_{B\dot{A}D\dot{C}F\dot{E}}(k_1,k_2)$ and $V_{B\dot{A}D\dot{C}F\dot{E}H\dot{G}}$
are the expressions for the three-gluon and four-gluon vertices, respectively. 
I further used the notation
\bq
 p_{i,j} & = & \sum\limits_{l=i}^j p_l.
\eq
The recursion starts with the current with one external leg, which is given by the polarisation vector:
\bq
J_1^{\dot{A}B}\left(p_1^\pm; q_1 \right) 
 & = & 
 \eps^{\dot{A}B}_\pm\left(p_1,q_1\right)
\eq

Similar recurrence relations can be written down for the quark- and antiquark currents, as well as the gluon
currents in full QCD.
The guiding principle is to follow the off-shell leg into the ``blob'', representing the sum of all diagrams,
and to sum on the r.h.s of the
recurrence relation over all vertices involving this off-shell leg and off-shell currents with less external
legs.


\section{The method}
\label{sect:method}

In this section I describe in detail the method for the automated computation of Born matrix elements in QCD
The matrix elements may or may not involve spin and/or colour correlations.

\subsection{Helicity amplitudes and spin correlations}

The program computes helicity amplitudes. For a given set of external momenta, each helicity amplitude 
evaluates to a complex number.
If no spin correlations are present, the matrix element is simply given as the squared modulus of the amplitude summed
over all helicity configurations.
In the dipole formalism, spin correlations are related to the splittings $g \rightarrow g g$ and $g \rightarrow q \bar{q}$.
In the original formulation of Catani and Seymour they are written as 
\bq
 {\cal A}^\ast_\mu\left(...,p_{(ij)},...\right) S^{\mu\nu} {\cal A}_\nu\left(...,p_{(ij)},...\right),
\eq
where ${\cal A}_\mu$ denotes the amplitude with the polarisation vector of the emitter gluon ${(ij)}$ amputated.
Furthermore, the spin correlation tensor is of the form
\bq
 S^{\mu\nu} & = & v^\mu v^\nu
\eq
and the vector $v^\mu$ satisfies
\bq
 v \cdot p_{(ij)} & = & 0.
\eq
Within the helicity formalism the spin correlation is evaluated as \cite{Weinzierl:1999yf}
\bq
 {\cal A}^\ast_\mu S^{\mu\nu} {\cal A}_\nu
  & = &
 \left| 
         E {\cal A}\left(...,p_{(ij)}^+,...\right) + E^\ast {\cal A}\left(...,p_{(ij)}^-,...\right) 
 \right|^2,
\eq
where ${\cal A}(...,p_{(ij)}^\pm,...)$ denotes the helicity amplitude, where the emitter gluon has ``$+$'', respectively ``$-$''
helicity.
$E$ is given by
\bq
\label{Efactor}
 E & = & \eps_-^\mu v_\mu = \frac{\l q+ | v | p_{(ij)}+ \r}{\sqrt{2} \left[ p_{(ij)} q \right]}.
\eq
In eq. (\ref{Efactor}) $q$ is as usual an arbitrary null reference momentum.

\subsection{Amplitudes with more than one quark-antiquark pair}

If more than one quark-antiquark pair is present, we have to sum over all quark permutations.
An amplitude with $n_q$ quark-antiquark pairs can be written as
\bq
\label{identical_quarks}
\lefteqn{
{\cal A}\left( \bar{q}_1, q_1, ..., \bar{q}_2, q_2, ..., \bar{q}_{n_q}, q_{n_q} \right)
 = } & & \nonumber \\
 & & 
 \sum\limits_{\sigma \in S(n_q)} \left( -1 \right)^{\sigma} 
 \left( \prod\limits_{j=1}^{n_q} \delta^{flav}_{\bar{q}_j q_{\sigma(j)} } \right)
     \hat{\cal A}\left( \bar{q}_1, q_{\sigma(1)}, ..., \bar{q}_2, q_{\sigma(2)}, ..., \bar{q}_{n_q}, q_{\sigma(n_q)} \right).
\eq
Here, $(-1)^\sigma$ equals $-1$ whenever the permutation is odd and equals $+1$ if the permutation is even.
In $\hat{\cal A}$ each external quark-antiquark pair $(\bar{q}_j, q_{\sigma(j)})$ is connected by a continuous fermion line.
The flavour factor $\delta^{flav}_{\bar{q}_j q_{\sigma(j)} }$ ensures that this combination is only taken into account, if
$\bar{q}_j$ and $ q_{\sigma(j)}$ have the same flavour.

\subsection{The colour structure}

\begin{figure}
\begin{center}
\begin{picture}(300,200)(0,0)
\Gluon(159,105)(211,135){4}{4}
\ArrowLine(219,112)(150,100)
\ArrowLine(150,100)(219,88)
\Gluon(159,95)(211,65){4}{4}
\Gluon(150,110)(150,170){4}{4}
\ArrowLine(150,100)(126,166)
\ArrowLine(105,154)(150,100)
\Gluon(141,105)(89,135){4}{4}
\Gluon(141,95)(89,65){4}{4}
\ArrowLine(150,100)(105,46)
\ArrowLine(126,34)(150,100)
\Gluon(150,90)(150,30){4}{4}
\GCirc(150,100){10}{0.5}
\Text(185,161)[]{$...$}
\Text(80,100)[]{$...$}
\Text(185,39)[]{$...$}
\Text(229,114)[]{$1$}
\Text(229,86)[]{$2$}
\Text(219,60)[]{$3$}
\Text(150,20)[]{$k$}
\Text(123,25)[]{$k+1$}
\Text(99,39)[]{$k+2$}
\Text(80,59)[]{$k+3$}
\Text(81,140)[]{$l$}
\Text(99,161)[]{$l+1$}
\Text(123,175)[]{$l+2$}
\Text(150,180)[]{$l+3$}
\Text(219,140)[]{$n$}
\end{picture}
\caption{\label{fig:cyclic_ordering}
The cyclic order of a partial amplitude. Without loss of generality we
can assume that quarks follow immediately antiquarks in the clockwise order.}
\end{center}
\end{figure}
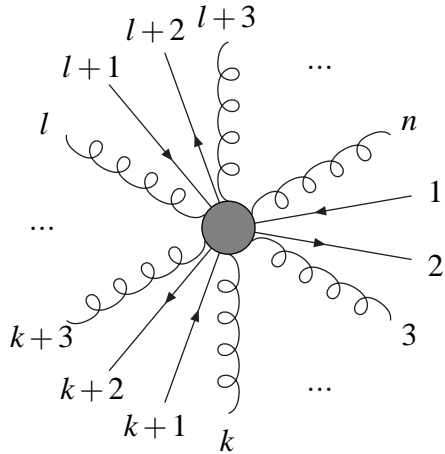

The amplitude $\hat{\cal A}$ is decomposed into colour factors and partial amplitudes:
\bq
\label{fixed_quark_flavours}
 \hat{\cal A} & = &
  \sum\limits_{i} c_i A_i
\eq
Each partial amplitude $A_i$ has a fixed cyclic ordering of the external legs. For Born graphs we can take 
this ordering such that
a quark follows immediately its corresponding antiquark in the clockwise orientation.
This is shown in fig. \ref{fig:cyclic_ordering}.
That is to say, that gluon are emitted from a quark line only to the right when following the fermion line arrow.
If a gluon would be emitted to the left, we could draw an equivalent diagram by flipping the off-shell current
attached to this gluon to the right of the fermion line.

All possible cyclic orderings are generated as follows: We assume that the amplitude has $n_g$ external gluons,
$n_q$  external quarks and therefore necessarily also $n_q$ external antiquarks. 
We first note, since a quark follows immediately its corresponding antiquark, we can treat an
adjacent $(\bar{q},q)$-pair as an external ``pseudo-leg'', which is permutated together.
The amplitude has therefore $n_g+n_q$ pseudo-legs. Then all possible cyclic orderings are obtained by
summing over all permutations of the pseudo-legs and factoring out the cyclic permutations, e.g.
each ordering corresponds to an element of
\bq
\label{sum_permutations}
 S(n_g+n_q) / Z(n_g+n_q).
\eq
This is equivalent to fixing the first external pseudo-leg and summing over all permutation of the remaining
$(n_g+n_q-1)$ external pseudo-legs.
Therefore there are
\bq
 (n_g+n_q-1)!
\eq
inequivalent cyclic orderings.

For the pure gluon amplitude ($n_q=0$) each cyclic ordering corresponds to one colour factor $c_i$.
The situation is different if quarks are present ($n_q \ne 0$).
This is related to the fact that the gluon propagator in an $SU(N)$ gauge theory can be written as
a propagator corresponding to an $U(N)$ gauge theory minus a part which subtracts out the additional 
$U(1)$ piece.
The kinematic parts of the $U(n)$ and the $U(1)$ pieces are the same:
\bq
P^{\dot{B}A\dot{D}C}(k) & = &
\begin{picture}(100,20)(0,5)
 \Gluon(20,10)(70,10){-5}{5}
 \Text(15,12)[r]{\footnotesize $\dot{B}A$}
 \Text(75,12)[l]{\footnotesize $\dot{D}C$}
\end{picture} 
 = 
\frac{i}{k^2}
 \left(
       - \eps^{\dot{B}\dot{D}} \eps^{AC}
 \right),
\eq
However, they differ by their colour structure:
\bq
U(N) : \;\;\;
\begin{picture}(85,20)(0,5)
 \ArrowLine(70,13)(20,13)
 \ArrowLine(20,7)(70,7)
 \Text(15,13)[rb]{\footnotesize $i$}
 \Text(15,7)[rt]{\footnotesize $j$}
 \Text(75,13)[lb]{\footnotesize $l$}
 \Text(75,7)[lt]{\footnotesize $k$}
\end{picture} 
 & = &
 \delta_{il} \delta_{kj},
 \nonumber \\
U(1) : \;\;\;
\begin{picture}(85,20)(0,5)
 \ArrowLine(25,13)(20,13)
 \Line(20,7)(25,7)
 \CArc(25,10)(3,-90,90)
 \Line(70,13)(65,13)
 \ArrowLine(65,7)(70,7)
 \CArc(65,10)(3,90,270)
 \DashLine(28,10)(62,10){5}
 \Text(15,13)[rb]{\footnotesize $i$}
 \Text(15,7)[rt]{\footnotesize $j$}
 \Text(75,13)[lb]{\footnotesize $l$}
 \Text(75,7)[lt]{\footnotesize $k$}
\end{picture} 
 & = &
 - \frac{1}{N} \delta_{ij} \delta_{kl}.
\eq
Note that each propagation of a $U(1)$ gluon is accompanied by a factor $(-1)/N$.
It can be shown that the $U(1)$ gluon couples only to quark lines
\cite{Maltoni:2002mq}.
Therefore an amplitude with $n_q$ quarks can contain up to $(n_q-1)$ gluons of type $U(1)$.
Each $U(1)$ gluon separates a Born amplitude into colour-disconnected pieces.
We define a colour cluster as a part of an amplitude, which is connected to the rest of the amplitude
only by an $U(1)$ gluon and which does not contain by itself any $U(1)$ gluon.
\begin{figure}
\begin{center}
\vspace*{-20mm}
\bq
\begin{picture}(130,100)(0,50)
 \Vertex(30,30){2}
 \Vertex(30,70){2}
 \Vertex(80,30){2}
 \Vertex(80,70){2}
 \ArrowLine(10,10)(30,30)
 \ArrowLine(30,30)(30,70)
 \ArrowLine(30,70)(10,90)
 \Gluon(30,30)(80,30){4}{3}
 \Gluon(30,70)(80,70){4}{3}
 \ArrowLine(110,80)(80,70)
 \ArrowLine(80,70)(110,60)
 \ArrowLine(110,40)(80,30)
 \ArrowLine(80,30)(110,20)
 \Text(5,10)[r]{\small $1$}
 \Text(5,90)[r]{\small $2$}
 \Text(115,80)[l]{\small $3$}
 \Text(115,60)[l]{\small $4$}
 \Text(115,40)[l]{\small $5$}
 \Text(115,20)[l]{\small $6$}
 \Text(55,75)[b]{\small $U(N)$}
 \Text(55,35)[b]{\small $U(1)$}
\end{picture}
 & \Longrightarrow & 
\begin{picture}(120,100)(0,50)
 \ArrowLine(10,10)(30,30)
 \ArrowLine(30,30)(30,68)
 \ArrowLine(30,72)(10,90)
 \DashLine(34,30)(76,30){4}
 \ArrowLine(80,72)(30,72)
 \ArrowLine(30,68)(80,68)
 \ArrowLine(110,80)(80,72)
 \ArrowLine(80,68)(110,60)
 \ArrowLine(110,40)(80,30)
 \ArrowLine(80,30)(110,20)
 \Text(5,10)[r]{\small $1$}
 \Text(5,90)[r]{\small $2$}
 \Text(115,80)[l]{\small $3$}
 \Text(115,60)[l]{\small $4$}
 \Text(115,40)[l]{\small $5$}
 \Text(115,20)[l]{\small $6$}
 \Text(55,75)[b]{\small $U(N)$}
 \Text(55,35)[b]{\small $U(1)$}
\end{picture}
\nonumber 
\eq
\vspace*{+10mm}
\caption{\label{fig:colour_cluster} An example for the decomposition into colour clusters.}
\end{center}
\end{figure}
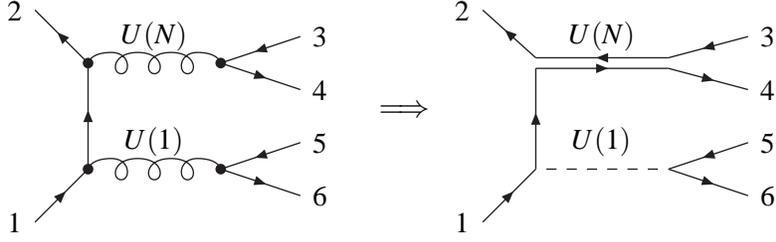
This concept is illustrated in fig. \ref{fig:colour_cluster}, which shows a diagram with three 
quark-antiquark pairs, one $U(N)$ gluon and one $U(1)$ gluon. This diagram has two colour
clusters, formed by the particles $(1,2,3,4)$ and $(5,6)$, and separated by the $U(1)$ gluon.
For an amplitude with $n_q$ quark-antiquark pairs one can have from $1$ to $n_q$ colour clusters.
From the cluster decomposition the colour structure can be read off easily.
The example in fig. \ref{fig:colour_cluster} contributes to the colour structure
\bq
 \left( - \frac{1}{N} \right) \left( \delta_{i_2 j_3} \delta_{i_4 j_1} \right) \left( \delta_{i_6 j_5} \right)
\eq
In general, given a colour cluster assignment, the corresponding colour factor $c_i$ is constructed as follows:
First of all, the colour factor factorizes into a product of the contributions from the individual colour clusters.
\bq 
 c_i & = &
 \left( - \frac{1}{N} \right)^{(n_{cluster}-1)}
 \times
 \prod\limits_{j=1}^{n_{cluster}} c_{i,j}.
\eq
$c_{i,j}$ is the colour factor corresponding to cluster $j$. For a cluster consisting only of gluons, $c_{i,j}$ is given
by
\bq
 g_1, g_2, ..., g_n: & & c_{i,j} = \delta_{i_1 j_2} \delta_{i_2 j_3} ... \delta_{i_{n-1} j_n} \delta_{i_n j_1}.
\eq
An antiquark-quark pair can be treated effectively as a single gluon. For example the colour factor associated to a 
colour cluster consisting of an quark-antiquark pair and $(n-2)$ gluons is given by
\bq
 \bar{q}_1, q_2, g_3, ..., g_n: & & c_{i,j} = \delta_{i_2 j_3} \delta_{i_3 j_4} ... \delta_{i_{n-1} j_n} \delta_{i_n j_1}.
\eq
As a further example we quote the colour factor for a cluster with two quark-antiquark pairs:
\bq
 \bar{q}_1, q_2, g_3, ..., \bar{q}_k, q_{k+1}, ..., g_n: & & 
  c_{i,j} = \delta_{i_2 j_3} \delta_{i_3 j_4} ... 
  \delta_{i_{k-1} j_k} \delta_{i_{k+1} j_{k+2}} ...
  \delta_{i_{n-1} j_n} \delta_{i_n j_1}.
\eq
The pattern should be clear.
The colour factor associated to a individual colour cluster is just a sequence of Kronecker $\delta$'s, corresponding to
the cyclic ordering of the legs belonging to this colour cluster.

It remains to derive a method, how all possible colour clusterings can be generated.
This is a combinatorial problem.
For a fixed cyclic ordering we can generate all possible colour clusterings as follows:
We first sum over the number of possible colour clusters.
Let $n_{cluster}$ be the number of colour clusters, where $n_{cluster}$ ranges from $1$ to $n_q$. 
For a fixed $n_{cluster}$ we then sum over all partitions of $(n_g+n_q)$ into $n_{cluster}$ pieces $n^{cluster}_j$, such that
\bq
 \sum\limits_{j=1}^{n_{cluster}} n^{cluster}_j & = & n_g+n_q.
\eq
For a partition we take into account the order, such that for example $(1,1,2)$, $(1,2,1)$ and $(2,1,1)$ are distinct partitions of
$4$.
$n^{cluster}_j$ gives the number of external pseudo-legs belonging to cluster $j$.
Obviously, an adjacent antiquark-quark pair has to belong to the same colour cluster,
therefore it is counted as one external pseudo-leg.
Finally, we have to sum over all possible starting points of the
colour clusters with respect to the cyclic ordering.
Here we observe that the members of a colour cluster need not be
adjacent in the cyclic ordering.
An example for a colour assignment in the cyclic ordering would be
\bq
 \underbrace{\left( \bar{q}, q \right), g, g}_{cluster 1}, 
 \underbrace{g, \left( \bar{q}, q \right), g}_{cluster 2}, 
 \underbrace{g, g, g, g}_{cluster 1}, 
 \underbrace{g, \left( \bar{q}, q \right), g, g}_{cluster 3}.
\eq
In this example, cluster $2$ is embedded in cluster $1$.
The summation over the starting points has to full-fill the following requirements: 
\begin{description}
\item{(i)} The external pseudo-leg $1$ belongs to colour cluster $1$.
\item{(ii)} The colour cluster $(j+1)$ starts after colour cluster $j$ for all $j>2$.
(Colour cluster $1$ may start at the end of the cyclic ordering.)
\item{(iii)} If the assignment of external pseudo-legs to colour cluster $j$ has been interrupted by
the starting of a new cluster $k$ ( with $k>j$ ), the assignment to cluster $j$ cannot be continued
until all members of cluster $k$ have been assigned.
\end{description}
Requirement (iii) ensures that we cannot have a sequence like cluster 1, cluster 2, cluster 1, cluster 2.
The assignment of the external pseudo-legs to colour clusters is now done as follows:
Let
\bq 
\label{m-tuple}
 \left( m_2^{start}, m_3^{start}, ..., m_{n_{cluster}}^{start} \right)
\eq
be an $(n_{cluster}-1)$-tuple, such that
\bq
\label{constraint_1}
 m_{j}^{start} \le m_{j+1}^{start}
\eq
and 
\bq
\label{constraint_2}
 1 \le m_{j}^{start} \le 2 - j + \sum\limits_{k=1}^{j-1} n_k^{cluster}
\eq
Then 
\bq
 n_j^{start} & = & m_j^{start} + j - 1
\eq
defines the starting point of cluster $j$ for $j=2,...,n_{cluster}$.
The starting points $n_j^{start}$ together with the rules (i) and (iii) define uniquely the assignment of the
external pseudo-legs to the colour clusters.
Summing over all $(n_{cluster}-1)$-tuples in eq. (\ref{m-tuple}) subject to the constraints
(\ref{constraint_1}) and (\ref{constraint_2}) generates all possibilities with $n_{cluster}$ colour clusters, in which
colour cluster $j$ has $n_j^{cluster}$ external pseudo-legs.

Since each colour cluster couples to the rest of the amplitude through a
$U(1)$-gluon, it has to contain at least one quark-antiquark pair.
Therefore configurations, where a colour cluster does not contain a quark-antiquark pair are vetoed,
with the trivial exception of the pure gluon amplitude, which consists of one colour cluster and no
quark-antiquark pairs.
\\
\\
With the colour cluster decomposition and a method for the generation of all cluster decomposition at hand,
I now turn back to the computation of the amplitude squared.
From eq. (\ref{identical_quarks}) and eq. (\ref{fixed_quark_flavours})
it is clear that we can write any amplitude in the form
\bq
\label{full_ampl}
 {\cal A} & = &
  \sum\limits_{i} c_i A_i,
\eq
where the $c_i$'s are the colour factors and the $A_i$'s are the partial amplitudes which contain
the kinematical information.
In squaring the amplitude we obtain
\bq
\label{ampl_squared}
 \left| {\cal A} \right|^2 & = & 
  \sum\limits_{i,j} A_i \left( c_i P c_j^\dagger \right) A_j^\ast.
\eq
The colour projector is given as a product with one factor for each external particle:
\bq
\label{colour_projector}
 P & = & \prod\limits_{k=1}^{n_g+2n_q} P_k,
\eq
where the individual colour projectors for a quark, antiquark and a gluon are:
\bq
\label{colour_projector_detailed}
 P_q = \delta_{\bar{i} i},
 \;\;\;
 P_{\bar{q}} = \delta_{j \bar{j}},
 \;\;\;
 P_g = \delta_{\bar{i}{i}} \delta_{j\bar{j}} - \frac{1}{N} \delta_{\bar{i}\bar{j}} \delta_{ji}.
\eq
The only non-trivial piece is given by the colour projector for the external gluons, which is a consequence of
the double-line notation.
Note that
\bq
 M_{ij} & = & \left( c_i P c_j^\dagger \right)
\eq 
defines a matrix, which is independent of the four-momenta of the particles.
Therefore this matrix can be calculated at the initialisation phase of the program.
As each entry is given as a contraction of Kronecker $\delta$'s, this can be done easily symbolically with the rules
\bq
 \delta_{ij} \delta_{jk} = \delta_{ik}, & & \delta_{ii} = N.
\eq
The program uses the C++ library ``GiNaC'' for this task.
In appendix (\ref{appendix:ginac}) I give a small example program.
The resulting expression is a function of $N$, and after substituting $N=3$ the result can be converted to
a double precision number.
Note that run-time performance is not an issue here, since this calculation occurs only at the initialisation
phase of the program.
To obtain the amplitude squared, the matrix $M_{ij}$ is first calculated at the initialisation phase and stored
in memory.
Then for each momentum configuration the vector of partial amplitudes $\vec{A} = (A_1,A_2,...)$ is computed.
The amplitude squared is then given by
\bq
 \left| {\cal A} \right|^2 & = & 
 \vec{A} \; M \; \vec{A}^\dagger.
\eq
The inclusion of colour-correlations is rather straightforward. 
To include colour-correlation between particles $a$ and $b$, one replaces
$P_a$ and $P_b$ in eq. (\ref{colour_projector}) by the appropriate colour-correlation operator.
For example, the colour-correlation operator ${\bf T}_q \cdot {\bf T}_{\bar{q}}$ for a quark-antiquark pair reads
\bq
\begin{picture}(100,35)(0,55)
\Vertex(50,40){2}
\Vertex(50,80){2}
\Gluon(50,40)(50,80){3}{6}
\ArrowLine(50,40)(80,40)
\ArrowLine(20,40)(50,40)
\ArrowLine(80,80)(50,80)
\ArrowLine(50,80)(20,80)
\Text(18,80)[r]{\small $\bar{i}_1$}
\Text(18,40)[r]{\small $\bar{j}_2$}
\Text(82,80)[l]{\small $i_1$}
\Text(82,40)[l]{\small $j_2$}
\end{picture}
 & = &
 - \frac{1}{2} 
 \left( 
  \delta_{\bar{i}_1 \bar{j}_2} \delta_{j_2 i_1}
  - \frac{1}{N} \delta_{\bar{i}_1 i_1} \delta_{j_2 \bar{j}_2}
 \right).
 \\ \nonumber
\eq
A complete list of all relevant colour-correlation operators can be found in the appendix (\ref{appendix:colour_operators}).
The corresponding matrices $M_{ij}$ depend now on $a$ and $b$, but are still independent of the four-momenta of the 
particles.
Therefore they can be computed at the initialisation phase of the program.
For a matrix element with $n=n_g+2n_q$ external particles, there are 
\bq  
 \frac{1}{2} n (n-1)
\eq
possibilities of choosing the colour-correlated partons $a$ and $b$. Therefore the initialisation phase
of the program
computes and stores $n(n-1)/2$ different colour matrices $M_{ij}$. For realistic values of $n$,
say $n < 9$, the CPU time and memory requirements for this task are rather modest.

\subsection{The partial amplitudes}

It remains to discuss how the partial amplitudes $A_i$, entering
eq. (\ref{full_ampl}) and eq. (\ref{ampl_squared}) are computed.
This is done with the help of off-shell currents and recurrence relations.
Compared to the introductory discussion of the pure gluonic off-shell current
in section \ref{subsect:recurrence} there are additional complications:
First of all, a rather trivial extension is given by the fact, that in full QCD
we have to allow for the possibility of multiple quark-antiquark pairs.
Secondly, and more important is the fact that the recurrence relations have to respect
the decomposition of a partial amplitude into colour clusters.
The algorithm is summarised as follows:
\begin{description}
\item{(i)} We consider coupled recurrence relations for the off-shell currents corresponding to
an $U(N)$-gluon, an $U(1)$-gluon, quarks and antiquarks.
Note that the $U(N)$-gluon and the $U(1)$-gluon are treated separately, as the latter
couples only to quarks.
It is also convenient to distinguish the off-shell currents for the quarks and antiquarks,
depending on whether the quarks are massive or massless. In the latter case specialised 
(and faster) routines can be used, since only helicity conserving interactions enter.
\item{(ii)} All recurrence relation express an off-shell current of type $A$ as a sum over off-shell
currents with fewer legs, which are combined through the basic three- or four-valent vertices of the theory.
The recurrence relations takes into account all possible interaction vertices, which contain $A$.
Note that the off-shell currents, which enter the r.h.s. of the recurrence relation need not be of
type $A$.
For the example, the recurrence relation for an $U(N)$-gluon involves the quark-antiquark-gluon vertex
and therefore the off-shell currents for a quark and an antiquark.
In general, the recurrence relations yield a coupled system of equations.
\item{(iii)} The off-shell parton for the quark-current, the antiquark-current and the $U(N)$-gluon-current belongs
to a specific colour cluster $a$.
The recurrence relation splits the off-shell current with $n$ external legs into off-shell currents with less
external legs.
This splitting has to respect the following selection rules:
\begin{description}
\item{-} For the $U(N)$-current, the off-shell current attached through the three- and four-gluon vertex have to contain
at least one parton belonging to colour cluster $a$.
In the off-shell quark- and antiquark-current, which are attached through the gluon-quark-antiquark vertex
to the $U(N)$-gluon current, the off-shell quark- and antiquark-lines have to belong to colour cluster $a$.
\item{-} For the off-shell quark current, the sub-current attached through an $U(N)$-gluon must contain at least
one parton belonging to colour cluster $a$. On the other hand, the sub-current attached through an $U(1)$-gluon
may not contain any parton of colour cluster $a$. Similar considerations apply to the antiquark current.
\item{-} Finally, the off-shell $U(1)$-current is rather simple and the recurrence relation involves an
quark- and an antiquark-current, whose off-shell legs necessarily belong to the same colour cluster.
\end{description}
\item (iv) As a further selection rule we have to veto configuration, 
where the off-shell current is divided into sub-currents between leg $j$ and $(j+1)$, in the case where these two legs belong to the
same colour cluster $b$, which is different from the colour cluster $a$ of the off-shell leg.
That is to say, that the recurrence relation where the off-shell leg belong to cluster $a$, cannot split
legs which belong to a different colour cluster $b$.
\end{description}

\subsection{The pure gluon amplitude}

In principle, the pure gluon amplitude can be treated with the methods discussed above. However the pure gluon amplitude
is a rather special case, which leads to many additional simplifications. Since it is known that pure gluonic processes
will contribute significantly to the cross section at the LHC, it is desirable to treat these processes separately with
optimised routines, taking into account the additional simplifications.
The simplifications are:
\begin{description}
\item{-} There is only one colour cluster and the colour decomposition is simply given by the $(n_g-1)!$ 
inequivalent cyclic orderings, as in eq.(\ref{colour_decomp_pure_gluon}).
\item{-} $U(1)$-gluons can be ignored and the recurrence relation for the partial amplitudes is given by
eq. (\ref{recurrence_pure_gluon}).
\item{-} In calculating the colour matrix $M_{ij}$, the colour projectors $P_g$ in eq. (\ref{colour_projector_detailed}) may
be replaced by
\bq
 P_g & \rightarrow &
 \delta_{\bar{i}{i}} \delta_{j\bar{j}}.
\eq
\end{description}

\subsection{QCD amplitudes with one electro-weak boson}

The methods discussed above require only minor modifications to include amplitudes with QCD partons and one
electro-weak boson. As these are relevant to electron-positron annihilation, electron-proton collisions or
$Z$-production at the LHC / Tevatron, these amplitudes have been implemented as well.
The amplitudes are computed by considering a recurrence relation, which couples the electro-weak current to an
off-shell quark current and an off-shell antiquark current.
   

\section{Numerical implementation}
\label{sect:num}

The algorithms discussed above have been implemented into a computer program.
This numerical program can compute Born matrix elements in QCD with spin- and colour-correlations.
To test the program I have first considered the case, where spin- and colour correlations are absent.
In this case one can compare the results with the ones from the program Madgraph.
I quote here the results of this comparison for processes with up to 
seven external particles.
The labelling of the momenta is 
\bq 
 p_1 p_2 \rightarrow p_3, p_4, ..., p_n.
\eq
$p_1$ and $p_2$ are the incoming momenta, $p_3$ to $p_n$ are the outgoing momenta.
For $2 \rightarrow 2$ processes I took the following set of momenta (in units of $\mbox{GeV}$):
{\small
\bq
  p_1 & = & (45.0, 0.0, 0.0, -45.0),
  \nonumber \\
  p_2 & = & (45.0, 0.0, 0.0, 45.0),
  \nonumber \\
  p_3 & = & (45.0, -20.8997, -29.6778, 26.5976),
  \nonumber \\
  p_4 & = & (45.0, 20.8997, 29.6778, -26.5976).
\eq
} 
The same initial state momenta $p_1$ and $p_2$ are used for all other processes.
The final state momenta for the $2 \rightarrow 3$ processes were chosen as
{\small
\bq
  p_3 & = & (41.8145, -9.20663, -26.7503, 30.7914),
  \nonumber \\
  p_4 & = & (17.3829, 12.8067, 10.7712, -4.70487),
  \nonumber \\
  p_5 & = & (30.8026, -3.6001, 15.9791, -26.0865).
\eq
} 
For $2 \rightarrow 4$ processes I used
{\small
\bq
  p_3 & = & (29.9152, -18.1846, -8.69254, 22.1061),
  \nonumber \\
  p_4 & = & (9.82719, 4.07529, 8.79524, -1.61538),
  \nonumber \\
  p_5 & = & (22.171, -9.26417, 14.187, -14.2988),
  \nonumber \\
  p_6 & = & (28.0866, 23.3735, -14.2898, -6.19197).
\eq
} 
Finally, for $2 \rightarrow 5$ processes I used
{\small
\bq
  p_3 & = & (20.165, -13.0392, 0.0298292, 15.3819),
  \nonumber \\
  p_4 & = & (9.60811, 2.4114, 9.15728, -1.6264),
  \nonumber \\
  p_5 & = & (20.5589, -7.64505, 15.4771, -11.166),
  \nonumber \\
  p_6 & = & (18.087, 17.056, -3.25968, -5.06046),
  \nonumber \\
  p_7 & = & (21.581, 1.21688, -21.4045, 2.47093).
\eq
} 
The strong coupling constant was taken to be $\alpha_s = 0.118$.
For this comparison, all quark masses have been set to zero. The flavour labels serve only to distinguish identical quarks from
non-identical quarks.
\begin{table}
\begin{center}
\begin{tabular}{|c|r|r|}
\hline
 Process & this work & Madgraph \\
\hline
 $ g g \rightarrow g g $                         & 56203.4 & 56203.2 \\
 $ g \bar{d} \rightarrow \bar{d} g $             & 8436.64 & 8436.62 \\
 $ \bar{u} \bar{d} \rightarrow \bar{d} \bar{u} $ & 1374.01 & 1374.01 \\
 $ \bar{d} \bar{d} \rightarrow \bar{d} \bar{d} $ & 1287.74 & 1287.74 \\
\hline
 $ g g \rightarrow g g g $                         & 21269.2 & 21269.3 \\
 $ g \bar{d} \rightarrow \bar{d} g g $             & 3222.01 & 3222.02 \\
 $ \bar{u} \bar{d} \rightarrow \bar{d} \bar{u} g $ & 56.459  & 56.4591 \\
 $ \bar{d} \bar{d} \rightarrow \bar{d} \bar{d} g $ & 53.2424 & 53.2425 \\
\hline
 $ g g \rightarrow g g g g $                                & 1354.24   & 1354.22 \\
 $ g \bar{d} \rightarrow \bar{d} g g g $                    & 138.691   & 138.689 \\
 $ \bar{u} \bar{d} \rightarrow \bar{d} \bar{u} g g $        & 0.975563  & 0.975546 \\
 $ \bar{d} \bar{d} \rightarrow \bar{d} \bar{d} g g $        & 0.902231  & 0.902215 \\
 $ \bar{u} \bar{d} \rightarrow \bar{d} \bar{u} \bar{s} s $  & 0.0116469 & 0.0116467 \\
 $ \bar{u} \bar{d} \rightarrow \bar{d} \bar{u} \bar{u} u $  & 0.0524928 & 0.0524927 \\
 $ \bar{d} \bar{d} \rightarrow \bar{d} \bar{d} \bar{d} d $  & 0.0583822 & 0.0583821 \\
\hline
 $ \bar{u} \bar{d} \rightarrow \bar{d} \bar{u} \bar{s} g s $ & 0.000453678 & 0.000453671 \\
 $ \bar{u} \bar{d} \rightarrow \bar{d} \bar{u} \bar{u} g u $ & 0.00202449  & 0.00202446 \\
\hline
\end{tabular}
\caption{\label{table:comparison}
Comparison of our program with Madgraph for various matrix elements with up to seven external particles.
}
\end{center}
\end{table}
Table \ref{table:comparison} shows the comparison of our program with Madgraph for the computation
of the matrix elements corresponding to the indicated processes.
The results do not contain any averaging over the colour degrees of freedom for the initial-state particles,
nor do they contain symmetry factors for the final-state particles.
As can be seen from the table, the agreement is satisfactory.

To check spin- and colour-correlations I have compared the program with existing NLO codes
for $e^+ e^- \rightarrow 4 \;\mbox{jets}$ \cite{Weinzierl:1999yf} and $p p \rightarrow t \bar{t} g$ \cite{Brandenburg:2004fw}.
\begin{table}
\begin{center}
\begin{tabular}{|l|rrrrr|}
\hline
 $n$  & 4 & 5 & 6 & 7 & 8\\
\hline
 time for $\; |{\cal A}(g_1,...,g_n)|^2$                           & 0.0006  & 0.009  & 0.18  & 4    &  127 \\
 time for $\; |{\cal A}(\bar{q}, q, g_3,...,g_n)|^2$               & 0.0004  & 0.003  & 0.05  & 0.6  &  14 \\
 time for $\; |{\cal A}(\bar{q}, q, \bar{q}', q', g_5,...,g_n)|^2$ & 0.0002  & 0.002  & 0.02  & 0.4  &  8 \\
\hline
\end{tabular}
\caption{\label{table:timing}
CPU time in seconds for the computation of some matrix elements summed over all helicities and colours on
a standard PC (Pentium IV with 2 GHz).
The examples consists of the $n$ gluon amplitudes, the amplitudes with an $\bar{q},q$-pair and $(n-2)$ gluons
and the amplitudes with two distinct $\bar{q},q$-pairs and $(n-4)$ gluons.
}
\end{center}
\end{table}

Table \ref{table:timing} gives an indication for the CPU time needed to evaluate matrix elements of increasing
complexity.
It gives the CPU time needed for the computation of the matrix elements, summed over all colours and spins, 
corresponding to the following cases:
The amplitude ${\cal A}(g_1,...,g_n)$ with $n$ gluons, 
the amplitude ${\cal A}(\bar{q}, q, g_3,...,g_n)$ with an $\bar{q},q$-pair and $(n-2)$ gluons
and the amplitude ${\cal A}(\bar{q}, q, \bar{q}', q', g_5,...,g_n)$ with two distinct $\bar{q},q$-pairs and $(n-4)$ gluons.


\section{Conclusions and outlook}
\label{sect:concl}

In this paper I discussed an algorithm for the automated computation of spin- and colour-correlated
Born matrix elements in QCD.
These matrix elements are needed for NLO calculations in combination with the subtraction method.
I implemented the algorithm into a computer program. The program handles QCD amplitudes with massless
and/or massive quarks.
In addition, I have implemented the extension to QCD amplitudes with one additional electro-weak boson.

The methods presented here are part of a larger project for the automated computation of observables
at next-to-leading order for LHC physics.
The remaining missing piece is the automated computation of the interference term of the one-loop amplitude
with the Born amplitude.
In a previous publication, we already reported on the automated computation of the one-loop integrals
entering the one-loop amplitude \cite{vanHameren:2005ed}.
Work on the automated computation of the interference term is in progress.

\subsection*{Acknowledgements}

I would like to thank Peter Uwer for useful discussions and 
for the comparison of the subtraction terms for $p p \rightarrow t \bar{t} g$.


\begin{appendix}

\section{Feynman rules}
\label{appendix:feynman}

In this appendix I summarise the colour-ordered Feynman rules. I extract from each
formula the coupling constant and split the remainder into a colour
part and a kinematical part.
 
\subsection{Propagators, polarisation vectors and polarisation sums}

\subsubsection*{Gluon propagator}

In Feynman gauge, the gluon propagator is given by $-i g^{\mu\nu} \delta^{ab}/k^2$. 
Contraction of the kinematical part $-i g^{\mu\nu}/k^2$
with $(1/2) \bar{\sigma}^{\mu\dot{B}A} \bar{\sigma}^{\nu\dot{D}C}$ yields:
\bq
P^{\dot{B}A\dot{D}C}(k) & = &
\begin{picture}(100,20)(0,5)
 \Gluon(20,10)(70,10){-5}{5}
 \Text(15,12)[r]{\footnotesize $\dot{B}A$}
 \Text(75,12)[l]{\footnotesize $\dot{D}C$}
\end{picture} 
 = 
\frac{i}{k^2}
 \left(
       - \eps^{\dot{B}\dot{D}} \eps^{AC}
 \right)
\eq
The colour factor $\delta^{ab}$
is contracted within the double-line notation with $\sqrt{2} T_{ij}^a \sqrt{2} T_{kl}^b$:
\bq
\sqrt{2} T_{ij}^a \; \delta^{ab} \; \sqrt{2} T_{kl}^b & = & 
   \delta_{il} \delta_{kj} - \frac{1}{N} \delta_{ij} \delta_{kl}
\eq
The colour structure is split into two pieces. The first piece
$\delta_{il} \delta_{kj}$ corresponds to the propagation of a $U(N)$ gluon, whereas the second piece
$-\delta_{ij} \delta_{kl}/N$ subtracts out the additional $U(1)$ gluon.
Schematically we have
\bq
\begin{picture}(85,20)(0,5)
 \ArrowLine(70,13)(20,13)
 \ArrowLine(20,7)(70,7)
 \Text(15,13)[rb]{\footnotesize $i$}
 \Text(15,7)[rt]{\footnotesize $j$}
 \Text(75,13)[lb]{\footnotesize $l$}
 \Text(75,7)[lt]{\footnotesize $k$}
\end{picture} 
 & = &
 \delta_{il} \delta_{kj},
 \nonumber \\
\begin{picture}(85,20)(0,5)
 \ArrowLine(25,13)(20,13)
 \Line(20,7)(25,7)
 \CArc(25,10)(3,-90,90)
 \Line(70,13)(65,13)
 \ArrowLine(65,7)(70,7)
 \CArc(65,10)(3,90,270)
 \DashLine(28,10)(62,10){5}
 \Text(15,13)[rb]{\footnotesize $i$}
 \Text(15,7)[rt]{\footnotesize $j$}
 \Text(75,13)[lb]{\footnotesize $l$}
 \Text(75,7)[lt]{\footnotesize $k$}
\end{picture} 
 & = &
 - \frac{1}{N} \delta_{ij} \delta_{kl}.
\eq
Note that each propagation of a $U(1)$ gluon is accompanied by a factor $(-1)/N$.

\subsubsection*{Quark propagator}

The kinematical piece of the quark propagator reads:
\bq
\frac{i}{p\!\!\!/ - m}
\eq
The colour factor is simply
\bq
\begin{picture}(85,20)(0,5)
 \ArrowLine(70,10)(20,10)
 \Text(15,10)[rb]{\footnotesize $i$}
 \Text(75,10)[lb]{\footnotesize $j$}
\end{picture} 
 & = &
 \delta_{ij}.
\eq

\subsubsection*{Gluon polarisation vectors and colour projector}

The gluon polarisation vectors are given by
\bq
\eps^{\dot{A}B}_+(k,q) =  
 \frac{1}{\l q k \r} \; k^{\dot{A}} q^B,
 & &
\eps^{\dot{A}B}_-(k,q) =  
 \frac{1}{[ k q ]} \; q^{\dot{A}} k^B.
\eq
$k$ is the momentum of the gluon and $q$ is an arbitrary light-like reference momentum.
The dependence on $q$ drops out in gauge-invariant quantities.

Colour factor: 
In the conventional approach we sum for the squared matrix element for each gluon over all eight colour degrees of freedom.
In the double-line notation a factor $\sqrt{2} T_{ij}^a$ is moved at each end into the colour projector.
Therefore, the colour projector reads
\bq
\sqrt{2} T_{ij}^a \;\; \sqrt{2} T_{kl}^a & = & 
   \delta_{il} \delta_{kj} - \frac{1}{N} \delta_{ij} \delta_{kl}.
\eq

\subsection{Vertices}

\subsubsection*{Quark-gluon vertex}

The kinematical part of the quark-gluon vertex is given by
\bq
\begin{picture}(100,35)(0,55)
\Vertex(50,50){2}
\Gluon(50,50)(50,80){3}{4}
\ArrowLine(80,50)(50,50)
\ArrowLine(50,50)(20,50)
\Text(50,82)[b]{\small $C\dot{D}$}
\Text(20,48)[t]{\small $A$}
\Text(80,48)[t]{\small $\dot{B}$}
\end{picture}
 = 
-i \sqrt{2} \eps_{CA} \eps_{\dot{D}\dot{B}},
 & &
\begin{picture}(100,35)(0,55)
\Vertex(50,50){2}
\Gluon(50,50)(50,80){3}{4}
\ArrowLine(80,50)(50,50)
\ArrowLine(50,50)(20,50)
\Text(50,82)[b]{\small $C\dot{D}$}
\Text(20,52)[b]{\small $\dot{A}$}
\Text(80,52)[b]{\small $B$}
\end{picture}
 = 
-i \sqrt{2} \delta_C^{\;\;B} \delta_{\dot{D}}^{\;\;\dot{A}}.
 \\ \nonumber
\eq
The colour factor is given by
\bq
\begin{picture}(100,35)(0,55)
\ArrowLine(52,50)(52,80)
\ArrowLine(48,80)(48,50)
\ArrowLine(80,50)(52,50)
\ArrowLine(48,50)(20,50)
\Text(50,82)[b]{\small $l k$}
\Text(18,50)[r]{\small $i$}
\Text(82,50)[l]{\small $j$}
\end{picture}
 & =  &
\frac{1}{\sqrt{2}} \delta_{il} \delta_{kj}.
\eq
Here I neglected terms proportional to  $\delta_{kl}$, which vanish when contracted into the gluon
propagator.
 
\subsubsection*{Three gluon vertex}

The kinematical part of the three-gluon vertex is given by
\bq
\begin{picture}(100,35)(0,55)
\Vertex(50,50){2}
\Gluon(50,50)(50,80){3}{4}
\Gluon(50,50)(76,35){3}{4}
\Gluon(50,50)(24,35){3}{4}
\LongArrow(56,70)(56,80)
\LongArrow(67,47)(76,42)
\LongArrow(33,47)(24,42)
\Text(60,80)[lt]{$k_{1,A\dot{B}}$}
\Text(78,35)[lc]{$k_{2,C\dot{D}}$}
\Text(22,35)[rc]{$k_{3,E\dot{F}}$}
\end{picture}
 & = &
\frac{i}{\sqrt{2}} \left[
        \eps_{CE} \eps_{\dot{D}\dot{F}} \left( k_3 - k_2 \right)_{A\dot{B}}
      + \eps_{EA} \eps_{\dot{F}\dot{B}} \left( k_1 - k_3 \right)_{C\dot{D}}
      + \eps_{AC} \eps_{\dot{B}\dot{D}} \left( k_2 - k_1 \right)_{E\dot{F}}
 \right].
 \nonumber \\
\eq
The colour factor reads:
\bq
\begin{picture}(100,35)(0,55)
\ArrowLine(48,50)(48,80)
\ArrowLine(52,80)(52,50)
\ArrowLine(52,50)(78,35)
\ArrowLine(76,32)(50,47)
\ArrowLine(22,35)(48,50)
\ArrowLine(50,47)(24,32)
\Text(52,80)[b]{\small $i_1 j_1$}
\Text(80,30)[t]{\small $j_2 i_2$}
\Text(20,30)[t]{\small $j_3 i_3$}
\end{picture}
 & = &
\frac{1}{\sqrt{2}} \delta_{i_1 j_2} \delta_{i_2 j_3} \delta_{i_3 j_1}.
 \\ \nonumber
 \\ \nonumber
\eq

\subsubsection*{Four gluon vertex}

The kinematical part of the four-gluon vertex is given by
\bq
\begin{picture}(100,35)(0,55)
\Vertex(50,50){2}
\Gluon(50,50)(71,71){3}{4}
\Gluon(50,50)(71,29){3}{4}
\Gluon(50,50)(29,29){3}{4}
\Gluon(50,50)(29,71){3}{4}
\Text(72,72)[lb]{\small $A\dot{B}$}
\Text(72,28)[lt]{\small $C\dot{D}$}
\Text(28,28)[rt]{\small $E\dot{F}$}
\Text(28,72)[rb]{\small $G\dot{H}$}
\end{picture}
 & = &
2 i \left[
         2 \eps_{AE} \eps_{\dot{B}\dot{F}} \eps_{CG} \eps_{\dot{D}\dot{H}}
         - \eps_{AC} \eps_{\dot{B}\dot{D}} \eps_{EG} \eps_{\dot{F}\dot{H}}
         - \eps_{AG} \eps_{\dot{B}\dot{H}} \eps_{CE} \eps_{\dot{D}\dot{F}}
 \right].
 \\ \nonumber 
 \\ \nonumber 
\eq
The colour factor reads:
\bq
\begin{picture}(100,35)(0,55)
\ArrowLine(50,53)(71,74)
\ArrowLine(74,71)(53,50)
\ArrowLine(53,50)(74,29)
\ArrowLine(71,26)(50,47)
\ArrowLine(50,47)(29,26)
\ArrowLine(26,29)(47,50)
\ArrowLine(47,50)(26,71)
\ArrowLine(29,74)(50,53)
\Text(72,72)[lb]{\small $i_1 j_1$}
\Text(72,28)[lt]{\small $j_2 i_2$}
\Text(28,28)[rt]{\small $j_3 i_3$}
\Text(28,72)[rb]{\small $i_4 j_4$}
\end{picture}
 & = &
\frac{1}{2} \delta_{i_1 j_2} \delta_{i_2 j_3} \delta_{i_3 j_4} \delta_{i_4 j_1}.
 \\ \nonumber 
 \\ \nonumber
\eq


\section{Colour correlations}
\label{appendix:colour_operators}

In this appendix I list all colour-correlation operators ${\bf T}_a \cdot {\bf T}_b$ between two partons in the 
double line notation.
\bq
 {\cal A}^\ast\left(  ... a, ..., b, ... \right) \left( {\bf T}_a \cdot {\bf T}_b \right) {\cal A}\left(  ... a, ..., b, ... \right).
\eq
Their action between amplitudes is defined in eq. (\ref{colour_charge_operator_final}) and eq. (\ref{colour_charge_operator_initial}).
As we write all amplitudes in the colour-flow decomposition, we would like to know the action of these operators
in this basis. 
In the following I denote the colour indices of the amplitude ${\cal A}^\ast$ with barred indices, the colour indices of the
amplitude ${\cal A}$ with un-barred indices.

\subsubsection*{Quark-quark ( ${\bf T}_q \cdot {\bf T}_q$ )}

\bq
\begin{picture}(100,35)(0,55)
\Vertex(50,40){2}
\Vertex(50,80){2}
\Gluon(50,40)(50,80){3}{6}
\ArrowLine(80,40)(50,40)
\ArrowLine(50,40)(20,40)
\ArrowLine(80,80)(50,80)
\ArrowLine(50,80)(20,80)
\Text(18,80)[r]{\small $\bar{i}_1$}
\Text(18,40)[r]{\small $\bar{i}_2$}
\Text(82,80)[l]{\small $i_1$}
\Text(82,40)[l]{\small $i_2$}
\end{picture}
 & = &
 \frac{1}{2} 
 \left( 
  \delta_{\bar{i}_1 i_2} \delta_{\bar{i}_2 i_1}
  - \frac{1}{N} \delta_{\bar{i}_1 i_1} \delta_{\bar{i}_2 i_2}
 \right)
 \\ \nonumber
\eq

\subsubsection*{Quark-antiquark ( ${\bf T}_q \cdot {\bf T}_{\bar{q}}$ )}

\bq
\begin{picture}(100,35)(0,55)
\Vertex(50,40){2}
\Vertex(50,80){2}
\Gluon(50,40)(50,80){3}{6}
\ArrowLine(50,40)(80,40)
\ArrowLine(20,40)(50,40)
\ArrowLine(80,80)(50,80)
\ArrowLine(50,80)(20,80)
\Text(18,80)[r]{\small $\bar{i}_1$}
\Text(18,40)[r]{\small $\bar{j}_2$}
\Text(82,80)[l]{\small $i_1$}
\Text(82,40)[l]{\small $j_2$}
\end{picture}
 & = &
 - \frac{1}{2} 
 \left( 
  \delta_{\bar{i}_1 \bar{j}_2} \delta_{j_2 i_1}
  - \frac{1}{N} \delta_{\bar{i}_1 i_1} \delta_{j_2 \bar{j}_2}
 \right)
 \\ \nonumber
\eq

\subsubsection*{Antiquark-antiquark ( ${\bf T}_{\bar{q}} \cdot {\bf T}_{\bar{q}}$ )}

\bq
\begin{picture}(100,35)(0,55)
\Vertex(50,40){2}
\Vertex(50,80){2}
\Gluon(50,40)(50,80){3}{6}
\ArrowLine(50,40)(80,40)
\ArrowLine(20,40)(50,40)
\ArrowLine(50,80)(80,80)
\ArrowLine(20,80)(50,80)
\Text(18,80)[r]{\small $\bar{j}_1$}
\Text(18,40)[r]{\small $\bar{j}_2$}
\Text(82,80)[l]{\small $j_1$}
\Text(82,40)[l]{\small $j_2$}
\end{picture}
 & = &
 \frac{1}{2} 
 \left( 
  \delta_{j_1 \bar{j}_2} \delta_{j_2 \bar{j}_1}
  - \frac{1}{N} \delta_{j_1 \bar{j}_1} \delta_{j_2 \bar{j}_2}
 \right)
 \\ \nonumber
\eq

\subsubsection*{Quark-gluon ( ${\bf T}_q \cdot {\bf T}_g$ )}

\bq
\begin{picture}(100,35)(0,55)
\Vertex(50,40){2}
\Vertex(50,80){2}
\Gluon(50,40)(50,80){3}{6}
\Gluon(80,40)(50,40){3}{4}
\Gluon(50,40)(20,40){3}{4}
\ArrowLine(80,80)(50,80)
\ArrowLine(50,80)(20,80)
\Text(18,80)[r]{\small $\bar{i}_1$}
\Text(18,40)[r]{\small $\bar{i}_2, \bar{j}_2$}
\Text(82,80)[l]{\small $i_1$}
\Text(82,40)[l]{\small $i_2, j_2$}
\end{picture}
 & = &
 \frac{1}{2} 
 \left( 
  \delta_{\bar{i}_1 i_2} \delta_{\bar{i}_2 i_1} \delta_{j_2 \bar{j}_2}
  - \delta_{\bar{i}_1 \bar{j}_2} \delta_{j_2 i_1} \delta_{\bar{i}_2 i_2}
 \right)
 \\ \nonumber
\eq

\subsubsection*{Antiquark-gluon ( ${\bf T}_{\bar{q}} \cdot {\bf T}_g$ )}

\bq
\begin{picture}(100,35)(0,55)
\Vertex(50,40){2}
\Vertex(50,80){2}
\Gluon(50,40)(50,80){3}{6}
\Gluon(80,40)(50,40){3}{4}
\Gluon(50,40)(20,40){3}{4}
\ArrowLine(50,80)(80,80)
\ArrowLine(20,80)(50,80)
\Text(18,80)[r]{\small $\bar{j}_1$}
\Text(18,40)[r]{\small $\bar{i}_2, \bar{j}_2$}
\Text(82,80)[l]{\small $j_1$}
\Text(82,40)[l]{\small $i_2, j_2$}
\end{picture}
 & = &
 \frac{1}{2} 
 \left( 
  \delta_{j_1 \bar{j}_2} \delta_{j_2 \bar{j}_1} \delta_{\bar{i}_2 i_2}
  - \delta_{j_1 i_2} \delta_{\bar{i}_2 \bar{j}_1} \delta_{j_2 \bar{j}_2}
 \right)
 \\ \nonumber
\eq

\subsubsection*{Gluon-gluon ( ${\bf T}_g \cdot {\bf T}_g$ )}

\bq
\begin{picture}(100,35)(0,55)
\Vertex(50,40){2}
\Vertex(50,80){2}
\Gluon(50,40)(50,80){3}{6}
\Gluon(80,40)(50,40){3}{4}
\Gluon(50,40)(20,40){3}{4}
\Gluon(80,80)(50,80){3}{4}
\Gluon(50,80)(20,80){3}{4}
\Text(18,80)[r]{\small $\bar{i}_1, \bar{j}_1$}
\Text(18,40)[r]{\small $\bar{i}_2, \bar{j}_2$}
\Text(82,80)[l]{\small $i_1, j_1$}
\Text(82,40)[l]{\small $i_2, j_2$}
\end{picture}
 & = &
 \frac{1}{2} 
 \left( 
  \delta_{\bar{i}_1 i_1} \delta_{\bar{i}_2 i_2} \delta_{j_1 \bar{j}_2} \delta_{j_2 \bar{j}_1}
  - \delta_{\bar{i}_1 i_1} \delta_{j_2 \bar{j}_2} \delta_{j_1 i_2} \delta_{\bar{i}_2 \bar{j}_1}
 \right.
 \nonumber \\
 & & \left.
  - \delta_{j_1 \bar{j}_1} \delta_{\bar{i}_2 i_2} \delta_{\bar{i}_1 \bar{j}_2} \delta_{j_2 i_1}
  + \delta_{j_1 \bar{j}_1} \delta_{j_2 \bar{j}_2} \delta_{\bar{i}_1 i_2} \delta_{\bar{i}_2 i_1}
 \right)
\eq


\section{Details on the implementation}

In this appendix I provide some details on the implementation of the algorithm into a C++ program.
I will discuss in a small example how the colour algebra is performed. 
I will also give some hints on the implementation of
loops over multi-indices like permutation, partitions, etc..

\subsection{Colour algebra}
\label{appendix:ginac}

Below I show a small program, which defines the colour structures
\bq
 c_1 = \delta_{i_1 j_2} \delta_{i_2 j_1},
 & &
 c_1^\dagger = \delta_{j_2 i_1} \delta_{j_1 i_2},
\eq
and contracts them:
\bq
 c_{11} & = & c_1 c_1^\dagger.
\eq
The result is obviously $c_{11}=N^2$, which equals $9$ for $N=3$.
\begin{verbatim}
#include <iostream>
#include ``ginac/ginac.h''

int main()
{
 using namespace GiNaC;

 // number of colours
 int Nc = 3;

 // define colour indices
 ex i1 = idx( symbol("i1"), Nc );
 ex i2 = idx( symbol("i2"), Nc );

 ex j1 = idx( symbol("j1"), Nc );
 ex j2 = idx( symbol("j2"), Nc );

 // define colour structures
 ex c1 = delta_tensor(i1,j2)*delta_tensor(i2,j1);
 ex c1_conj = delta_tensor(j2,i1)*delta_tensor(j1,i2);

 // square it and contract indices
 ex c11 = c1_conj * c1;
 c11 = c11.simplify_indexed();
 
 // convert the result to a ``double'' variable
 double c_double = real(ex_to<numeric>( c11 )).to_double();

 std::cout << ``result = `` << c_double << std::endl;

 return 0;
}
\end{verbatim}

\subsection{Summing over multi-indices}

The algorithms involves the summation over multi-indices. A rather simple example for a multi-index
would be a $k$-tuple $(i_0,i_1,...,i_{k-1})$ where each entry can take values from $0$ to $N-1$.
Other examples are the sum over permutations of $k$ elements as in eq. (\ref{sum_permutations}) or the multi-index in
eq. (\ref{m-tuple}).
To make the code readable it is desirable to write the loop as
\begin{verbatim}
{
 int N = 7;
 int k = 3;

 multi_index i_multi(N,k);

 for( i_multi.init(); !i_multi.overflow(); i_multi++)
   {
    // can use i_multi[0], i_multi[1], etc. here
   }
}
\end{verbatim}
and to hide the details on how the multi-index is increased into a separate class.
A possible header file for the class \v/multi_index/ could look as follows:
\begin{verbatim}
class multi_index {

  public :  
    multi_index(size_t N, size_t k);

      // functions 
    multi_index & init(void);           // initialization
    bool overflow(void) const;          // returns overflow flag
    multi_index & operator++ (int);     // postfix increment
    size_t operator[](size_t i) const;  // subscripting
 
      // member variables :
  protected : 
    size_t N;
    std::vector<size_t> v;
    bool flag_overflow;
};
\end{verbatim}
This class contains a method \v/init/ to initialise the multi-index to the first value, an operator \v/++/ which increases
the multi-index to the next value and method \v/overflow/, which returns true if all values have been run through.

\end{appendix}




\begin{thebibliography}{10}

\bibitem{Berends:1987me}
F.~A. Berends and W.~T. Giele,
\newblock Nucl. Phys. {\bf B306}, 759 (1988).

\bibitem{Berends:1989ie}
F.~A. Berends, W.~T. Giele, and H.~Kuijf,
\newblock Phys. Lett. {\bf B232}, 266 (1989).

\bibitem{Berends:1990ax}
F.~A. Berends, H.~Kuijf, B.~Tausk, and W.~T. Giele,
\newblock Nucl. Phys. {\bf B357}, 32 (1991).

\bibitem{Caravaglios:1995cd}
F.~Caravaglios and M.~Moretti,
\newblock Phys. Lett. {\bf B358}, 332 (1995), hep-ph/9507237.

\bibitem{Caravaglios:1998yr}
F.~Caravaglios, M.~L. Mangano, M.~Moretti, and R.~Pittau,
\newblock Nucl. Phys. {\bf B539}, 215 (1999), hep-ph/9807570.

\bibitem{Draggiotis:1998gr}
P.~Draggiotis, R.~H.~P. Kleiss, and C.~G. Papadopoulos,
\newblock Phys. Lett. {\bf B439}, 157 (1998), hep-ph/9807207.

\bibitem{Draggiotis:2002hm}
P.~D. Draggiotis, R.~H.~P. Kleiss, and C.~G. Papadopoulos,
\newblock Eur. Phys. J. {\bf C24}, 447 (2002), hep-ph/0202201.

\bibitem{Stelzer:1994ta}
T.~Stelzer and W.~F. Long,
\newblock Comput. Phys. Commun. {\bf 81}, 357 (1994), hep-ph/9401258.

\bibitem{Pukhov:1999gg}
A.~Pukhov {\em et~al.},
\newblock (1999), hep-ph/9908288.

\bibitem{Yuasa:1999rg}
F.~Yuasa {\em et~al.},
\newblock Prog. Theor. Phys. Suppl. {\bf 138}, 18 (2000), hep-ph/0007053.

\bibitem{Krauss:2001iv}
F.~Krauss, R.~Kuhn, and G.~Soff,
\newblock JHEP {\bf 02}, 044 (2002), hep-ph/0109036.

\bibitem{Kilgore:1996sq}
W.~B. Kilgore and W.~T. Giele,
\newblock Phys. Rev. {\bf D55}, 7183 (1997), hep-ph/9610433.

\bibitem{Nagy:2001fj}
Z.~Nagy,
\newblock Phys. Rev. Lett. {\bf 88}, 122003 (2002), hep-ph/0110315.

\bibitem{Nagy:2003tz}
Z.~Nagy,
\newblock Phys. Rev. {\bf D68}, 094002 (2003), hep-ph/0307268.

\bibitem{Campbell:2002tg}
J.~Campbell and R.~K. Ellis,
\newblock Phys. Rev. {\bf D65}, 113007 (2002), hep-ph/0202176.

\bibitem{Beenakker:2002nc}
W.~Beenakker {\em et~al.},
\newblock Nucl. Phys. {\bf B653}, 151 (2003), hep-ph/0211352.

\bibitem{Dawson:2003zu}
S.~Dawson, C.~Jackson, L.~H. Orr, L.~Reina, and D.~Wackeroth,
\newblock Phys. Rev. {\bf D68}, 034022 (2003), hep-ph/0305087.

\bibitem{DelDuca:2001eu}
V.~Del~Duca, W.~Kilgore, C.~Oleari, C.~Schmidt, and D.~Zeppenfeld,
\newblock Phys. Rev. Lett. {\bf 87}, 122001 (2001), hep-ph/0105129.

\bibitem{DelDuca:2001fn}
V.~Del~Duca, W.~Kilgore, C.~Oleari, C.~Schmidt, and D.~Zeppenfeld,
\newblock Nucl. Phys. {\bf B616}, 367 (2001), hep-ph/0108030.

\bibitem{Soper:1998ye}
D.~E. Soper,
\newblock Phys. Rev. Lett. {\bf 81}, 2638 (1998), hep-ph/9804454.

\bibitem{Soper:1999xk}
D.~E. Soper,
\newblock Phys. Rev. {\bf D62}, 014009 (2000), hep-ph/9910292.

\bibitem{Passarino:2001wv}
G.~Passarino,
\newblock Nucl. Phys. {\bf B619}, 257 (2001), hep-ph/0108252.

\bibitem{Ferroglia:2002mz}
A.~Ferroglia, M.~Passera, G.~Passarino, and S.~Uccirati,
\newblock Nucl. Phys. {\bf B650}, 162 (2003), hep-ph/0209219.

\bibitem{Nagy:2003qn}
Z.~Nagy and D.~E. Soper,
\newblock JHEP {\bf 09}, 055 (2003), hep-ph/0308127.

\bibitem{Denner:2002ii}
A.~Denner and S.~Dittmaier,
\newblock Nucl. Phys. {\bf B658}, 175 (2003), hep-ph/0212259.

\bibitem{Dittmaier:2003bc}
S.~Dittmaier,
\newblock Nucl. Phys. {\bf B675}, 447 (2003), hep-ph/0308246.

\bibitem{Giele:2004iy}
W.~T. Giele and E.~W.~N. Glover,
\newblock JHEP {\bf 04}, 029 (2004), hep-ph/0402152.

\bibitem{Ellis:2005zh}
R.~K. Ellis, W.~T. Giele, and G.~Zanderighi,
\newblock (2005), hep-ph/0508308.

\bibitem{delAguila:2004nf}
F.~del Aguila and R.~Pittau,
\newblock JHEP {\bf 07}, 017 (2004), hep-ph/0404120.

\bibitem{Pittau:2004bc}
R.~Pittau,
\newblock (2004), hep-ph/0406105.

\bibitem{vanHameren:2005ed}
A.~van Hameren, J.~Vollinga, and S.~Weinzierl,
\newblock Eur. Phys. J. {\bf C41}, 361 (2005), hep-ph/0502165.

\bibitem{Binoth:2002xh}
T.~Binoth, G.~Heinrich, and N.~Kauer,
\newblock Nucl. Phys. {\bf B654}, 277 (2003), hep-ph/0210023.

\bibitem{Binoth:2005ff}
T.~Binoth, J.~P. Guillet, G.~Heinrich, E.~Pilon, and C.~Schubert,
\newblock (2005), hep-ph/0504267.

\bibitem{vanHameren:2004wr}
A.~van Hameren and C.~G. Papadopoulos,
\newblock Acta Phys. Polon. {\bf B35}, 2601 (2004), hep-ph/0410189.

\bibitem{Giele:1992vf}
W.~T. Giele and E.~W.~N. Glover,
\newblock Phys. Rev. {\bf D46}, 1980 (1992).

\bibitem{Giele:1993dj}
W.~T. Giele, E.~W.~N. Glover, and D.~A. Kosower,
\newblock Nucl. Phys. {\bf B403}, 633 (1993), hep-ph/9302225.

\bibitem{Keller:1998tf}
S.~Keller and E.~Laenen,
\newblock Phys. Rev. {\bf D59}, 114004 (1999), hep-ph/9812415.

\bibitem{Frixione:1996ms}
S.~Frixione, Z.~Kunszt, and A.~Signer,
\newblock Nucl. Phys. {\bf B467}, 399 (1996), hep-ph/9512328.

\bibitem{Catani:1997vz}
S.~Catani and M.~H. Seymour,
\newblock Nucl. Phys. {\bf B485}, 291 (1997), hep-ph/9605323.

\bibitem{Catani:1997vzerr}
S.~Catani and M.~H. Seymour,
\newblock Nucl. Phys. {\bf B510}, 503 (1997),
\newblock Erratum.

\bibitem{Dittmaier:1999mb}
S.~Dittmaier,
\newblock Nucl. Phys. {\bf B565}, 69 (2000), hep-ph/9904440.

\bibitem{Phaf:2001gc}
L.~Phaf and S.~Weinzierl,
\newblock JHEP {\bf 04}, 006 (2001), hep-ph/0102207.

\bibitem{Catani:2002hc}
S.~Catani, S.~Dittmaier, M.~H. Seymour, and Z.~Trocsanyi,
\newblock Nucl. Phys. {\bf B627}, 189 (2002), hep-ph/0201036.

\bibitem{Bauer:2000cp}
C.~Bauer, A.~Frink, and R.~Kreckel,
\newblock J. Symbolic Computation {\bf 33}, 1 (2002), cs.sc/0004015.

\bibitem{Berends:1981rb}
F.~A. Berends, R.~Kleiss, P.~De~Causmaecker, R.~Gastmans, and T.~T. Wu,
\newblock Phys. Lett. {\bf B103}, 124 (1981).

\bibitem{DeCausmaecker:1982bg}
P.~De~Causmaecker, R.~Gastmans, W.~Troost, and T.~T. Wu,
\newblock Nucl. Phys. {\bf B206}, 53 (1982).

\bibitem{Gunion:1985vc}
J.~F. Gunion and Z.~Kunszt,
\newblock Phys. Lett. {\bf B161}, 333 (1985).

\bibitem{Xu:1987xb}
Z.~Xu, D.-H. Zhang, and L.~Chang,
\newblock Nucl. Phys. {\bf B291}, 392 (1987).

\bibitem{Gastmans:1990xh}
R.~Gastmans and T.~T. Wu,
\newblock Oxford, UK: Clarendon (1990) 648 p. (International series of
  monographs on physics, 80).

\bibitem{Cvitanovic:1980bu}
P.~Cvitanovic, P.~G. Lauwers, and P.~N. Scharbach,
\newblock Nucl. Phys. {\bf B186}, 165 (1981).

\bibitem{Berends:1987cv}
F.~A. Berends and W.~Giele,
\newblock Nucl. Phys. {\bf B294}, 700 (1987).

\bibitem{Mangano:1987xk}
M.~L. Mangano, S.~J. Parke, and Z.~Xu,
\newblock Nucl. Phys. {\bf B298}, 653 (1988).

\bibitem{Kosower:1987ic}
D.~Kosower, B.-H. Lee, and V.~P. Nair,
\newblock Phys. Lett. {\bf B201}, 85 (1988).

\bibitem{Bern:1990ux}
Z.~Bern and D.~A. Kosower,
\newblock Nucl. Phys. {\bf B362}, 389 (1991).

\bibitem{DelDuca:1999rs}
V.~Del~Duca, L.~J. Dixon, and F.~Maltoni,
\newblock Nucl. Phys. {\bf B571}, 51 (2000), hep-ph/9910563.

\bibitem{Maltoni:2002mq}
F.~Maltoni, K.~Paul, T.~Stelzer, and S.~Willenbrock,
\newblock Phys. Rev. {\bf D67}, 014026 (2003), hep-ph/0209271.

\bibitem{Kosower:1989xy}
D.~A. Kosower,
\newblock Nucl. Phys. {\bf B335}, 23 (1990).

\bibitem{Cachazo:2004kj}
F.~Cachazo, P.~Svrcek, and E.~Witten,
\newblock JHEP {\bf 09}, 006 (2004), hep-th/0403047.

\bibitem{Britto:2004ap}
R.~Britto, F.~Cachazo, and B.~Feng,
\newblock (2004), hep-th/0412308.

\bibitem{Bena:2004ry}
I.~Bena, Z.~Bern, and D.~A. Kosower,
\newblock (2004), hep-th/0406133.

\bibitem{Schwinn:2005pi}
C.~Schwinn and S.~Weinzierl,
\newblock JHEP {\bf 05}, 006 (2005), hep-th/0503015.

\bibitem{'tHooft:1973jz}
G.~'t~Hooft,
\newblock Nucl. Phys. {\bf B72}, 461 (1974).

\bibitem{vanderHeide:2000fx}
J.~van~der Heide, E.~Laenen, L.~Phaf, and S.~Weinzierl,
\newblock Phys. Rev. {\bf D62}, 074025 (2000), hep-ph/0003318.

\bibitem{Weinzierl:1999yf}
S.~Weinzierl and D.~A. Kosower,
\newblock Phys. Rev. {\bf D60}, 054028 (1999), hep-ph/9901277.

\bibitem{Brandenburg:2004fw}
A.~Brandenburg, S.~Dittmaier, P.~Uwer, and S.~Weinzierl,
\newblock Nucl. Phys. Proc. Suppl. {\bf 135}, 71 (2004), hep-ph/0408137.

\end{thebibliography}
\end{document}